\documentclass{article}

\usepackage{arxiv}
\usepackage[utf8]{inputenc} 
\usepackage[T1]{fontenc}    
\usepackage{hyperref}       
\usepackage{url}            
\usepackage{booktabs}       
\usepackage{amsfonts}       
\usepackage{nicefrac}       
\usepackage{microtype}      
\usepackage{lipsum}
\usepackage{graphicx}
\usepackage{amssymb}
\usepackage{amsmath,amsthm}
\usepackage{array}
\usepackage{subfig}
\usepackage{capt-of}
\usepackage{mathtools}
\usepackage{lineno}         
\usepackage[table]{xcolor}  
\usepackage{comment}

\usepackage[backend=biber, bibstyle=nejm, sorting=nyt, citestyle=authoryear-comp, uniquename=false, uniquelist=false, mincitenames=1, maxcitenames=2, minbibnames=6, maxbibnames=6]{biblatex}

\title{Genome-wide networks reveal emergence of epidemic strains of Salmonella Enteritidis}

\author{
  Adam J. ~Svahn \textsuperscript{a,b*} \And
  Sheryl L. ~Chang \textsuperscript{a} \And 
  Rebecca J. ~Rockett \textsuperscript{b,c} \And
  Oliver M. ~Cliff \textsuperscript{a,d} \And
  Qinning ~Wang \textsuperscript{c,e} \And
  Alicia ~Arnott \textsuperscript{c} \And
  Marc ~Ramsperger \textsuperscript{c} \And
  Tania C. ~Sorrell \textsuperscript{b,c} \And
  Vitali ~Sintchenko \textsuperscript{b,c,e} \And
  Mikhail ~Prokopenko \textsuperscript{a,b} \and
  \\
  \textsuperscript{a}Complex Systems Research Group, Faculty of Engineering \\
  The University of Sydney, Sydney, NSW, Australia \\
  \textsuperscript{b}The University of Sydney Institute of Infectious Diseases \\
   The University of Sydney, Westmead, NSW, Australia\\
  \textsuperscript{c}Centre for Infectious Diseases and Microbiology–Public Health \\
   Westmead Hospital, Westmead, New South Wales, Australia \\
  \textsuperscript{d}School of Physics, Faculty of Science \\
   The University of Sydney, Sydney, NSW, Australia \\
  \textsuperscript{e}NSW Enteric Reference Laboratory, Institute of Clinical Pathology and Medical Research \\
  NSW Health Pathology, Westmead, NSW, Australia \\
  \textsuperscript{*}Corresponding author, adam.svahn@sydney.edu.au \\
 }
\begin{document}

\keywords{Salmonella Enteritidis \and Foodborne pathogen \and genotype network}

\maketitle


\begin{abstract}
\subsection*{Objectives}
To enhance monitoring of high-burden foodborne pathogens, there is opportunity to combine pangenome data with network analysis.

\subsection*{Methods}
Salmonella \textit{enterica} subspecies Enterica serovar Enteritidis isolates were referred to the New South Wales (NSW) Enteric Reference Laboratory between August 2015 and December 2019 (1033 isolates in total), inclusive of a confirmed outbreak. All isolates underwent whole genome sequencing. Distances between genomes were quantified by in silico MLVA as well as core SNPs, which informed construction of undirected networks. Prevalence-centrality spaces were generated from the undirected networks.  Components on the undirected SNP network were considered alongside a phylogenetic tree representation.

\subsection*{Results}
Outbreak isolates were identifiable as distinct components on the MLVA and SNP networks. The MLVA network based centrality/prevalence space did not delineate the outbreak, whereas the outbreak was clearly delineated in the SNP network based centrality/prevalence space. Components on the undirected SNP network showed a high concordance to the SNP clusters based on phylogenetic analysis.

\subsection*{Conclusions}
Bacterial whole genome data in network based analysis can improve the resolution of population analysis. High concordance of network components and SNP clusters is promising for rapid population analyses of foodborne \textit{Salmonella} spp. due to the low overhead of network analysis.

\end{abstract}

\section{Introduction}
\label{intro}

\textit{Salmonella enterica} is estimated to be responsible for the largest foodborne disease burden globally\parencite{kirk_world_2015}. The most common pathology resulting from foodborne non-typhoidal Salmonella infection is salmonella enterocolitis, a disease which affects 78.5 million people world-wide every year with over 29,000 deaths globally (2010,\parencite{kirk_world_2015}).

In Australia, non-typhoidal Salmonella is responsible for approximately 40,000 foodborne infections annually, requiring an estimated 2100 hospitalisations and causing approximately 15 deaths \parencite{kirk_foodborne_2014}. Although this highly diverse bacterial species contains over 2600 different serovars, two non-typhoidal Salmonella serovars, \textit{S.} Enteritidis and \textit{S.} Typhimurium, predominate globally and in Australia \parencite{moffatt_salmonella_2016}. In the USA and UK, \textit{S.} Enteritidis is the predominant causal agent \parencite{martelli_salmonella_2012} whereas in Australia S. Typhimurium (STM) causes the majority of locally acquired infections \parencite{sotomayor_novel_2018}. In Australia, \textit{S.} Enteritidis infections have been observed in travellers returning from endemic areas \parencite{chousalkar_review_2018, moffatt_salmonella_2016}. Monitoring of poultry sources for \textit{Salmonella} spp. has indicated to date that \textit{S.} Enteritidis is not endemic in local commercial poultry flocks \parencite{chousalkar_salmonella_2017}. As a known opportunistic foodborne pathogen that may establish itself in a naive population and begin to cause locally acquired infections, the monitoring of overseas and locally acquired \textit{S.} Enteritidis is of high priority to public health \parencite{marais_improving_2019}.


Genomic public health surveillance of foodborne infections has been rapidly becoming the gold standard in disease control \parencite{allard_practical_2016, ashton_identification_2016}. In New South Wales (NSW), bacterial isolates obtained from presentations of \textit{S.} Enteritidis are subjected to routine whole genome sequencing (WGS) and each case undergoes public health follow up to determine the source of infections. The data for this study comprised WGS data for \textit{S.} Enteritidis isolates identified in NSW over the period of 2015-2019. Although WGS analysis enables unparalleled resolution that accurately identifies food-borne outbreaks, the bioinformatic analysis of this diverse species can be challenging. In Australia \textit{S.} Enteritidis genomes can be phylogenetically classified into three major lineages, with significant genomic diversity between each \parencite{graham_comparative_2018}. To maintain the high resolution needed to pinpoint foodborne outbreaks the reference genome used in the bioinformatic process must be closely related to the genomes of isolates associated with the outbreak. When novel or emerging strains are detected, determining or generating the ideal reference genome can be challenging and time consuming. Therefore we sought to apply novel network-based methods using WGS data to better predict high-risk emerging strains and more rapidly identify outbreaks.

In the present study, we first extended our network models \parencite{cliff_network_2019, cliff_inferring_2020} to utilise WGS data and successfully identified the emergence of a strain responsible for a community outbreak using the networks inferred for the \textit{S.} Enteritidis population. We then demonstrated that epidemic strains can be discriminated within the centrality-prevalence space defined for the constructed networks, while network components represent a method of suitably partitioning the population. The findings suggested that network-based analytic methods can augment epidemiological characterisation and complement conventional phylogenetic clustering methods, with particular utility and application to sporadic and heterogeneous data typical of epidemiological monitoring.

\section{Results}
\label{results}

\subsection{Dynamics of \textit{S}. Enteritidis infections over the study period}

The total number of isolates over the period was 1033 which was reduced to 897 after removing assemblies which did not meet the quality criteria as outlined in the supplementary methods. The resulting core genome contained 4084 core genes resulting in a 3769845-bp core genome alignment. The majority of isolates clustered with the predominate lineage I, the clade containing four divergent lineage I isolates was collapsed to improve visual representation of the phylogeny all three representations of the phylogeny are contained in Supplementary figure \ref{fig:Max_likelihood_phylo}). 

\begin{figure}[h]
    \centering
    \includegraphics[width=\textwidth, trim={1cm 0cm 1cm 0cm},clip]{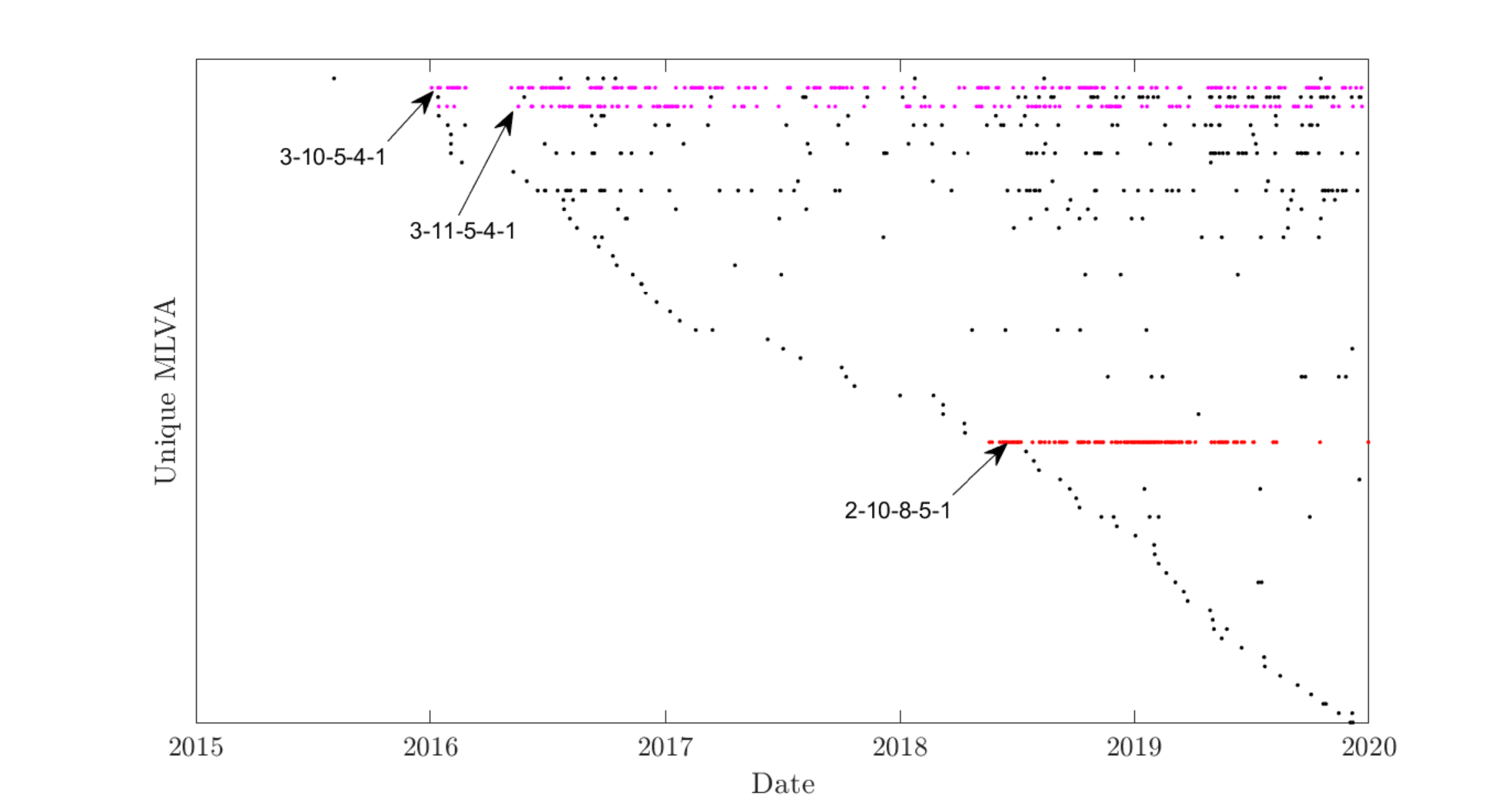}
    \caption{Unique MLVA profiles isolated between 4th August 2015 and 2nd January 2020. Number of unique profiles: 72. Each row is a unique MLVA profile and each dot represents an individual isolate. Magenta colour represents isolates of MLVA profiles which appeared consistently throughout the data set. Red colour represents the isolates associated with the outbreak commencing in May 2018.}
    \label{fig:time_MLVA}
\end{figure}

\begin{figure}[h]
    \centering
    \includegraphics[width=\textwidth]{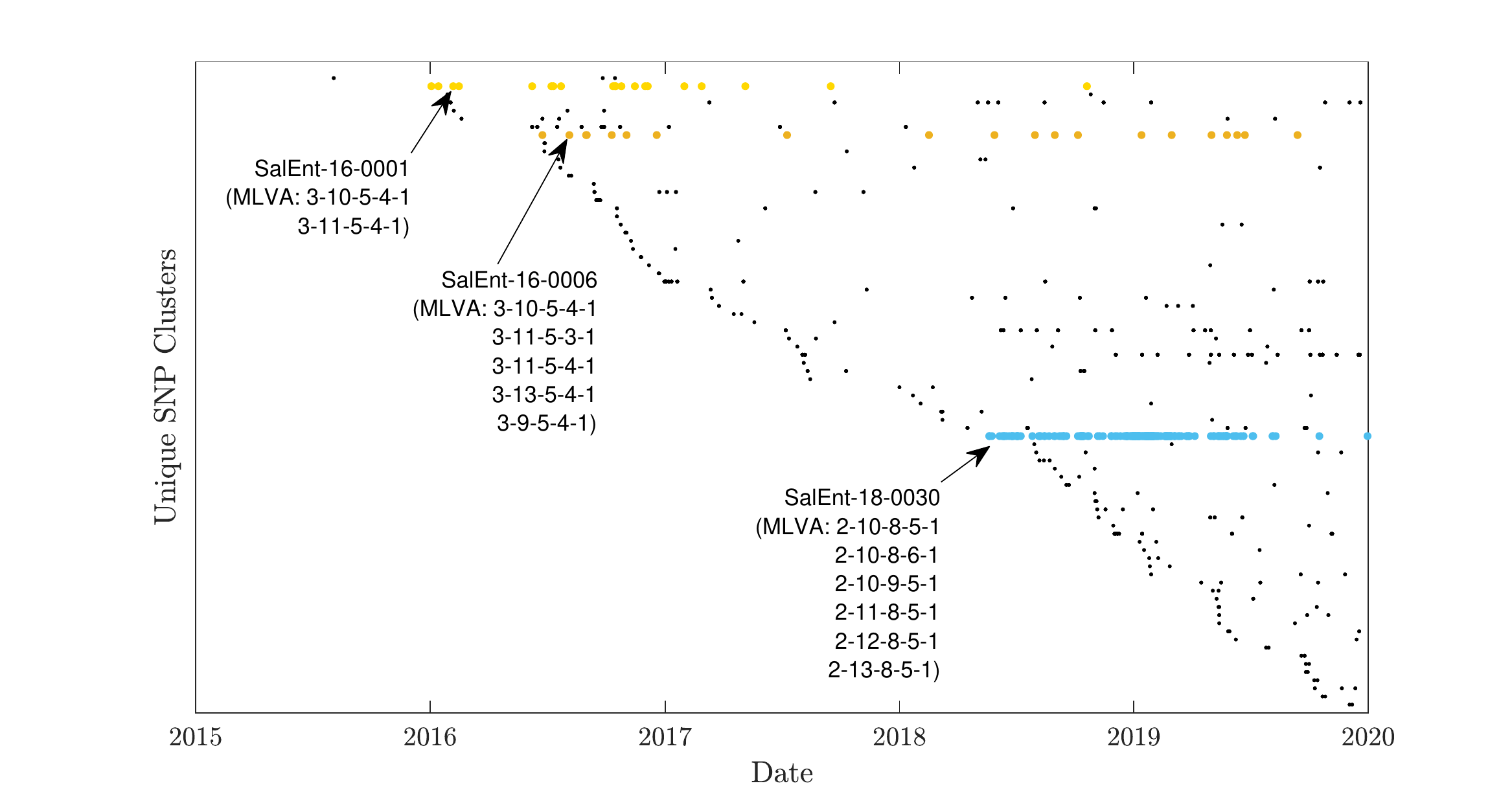}
    \caption{Unique SNP clusters from the isolated MLVA profiles. Number of unique profiles: 106. SNP clusters SalEnt-16-0001 (yellow) and SalEnt-16-0006 (orange) are highlighted as these contain the isolates belonging to MLVA profiles 2-10-5-4-1 and 2-11-5-4-1 which were recurring throughout the data set, as shown in Fig.~\ref{fig:time_MLVA}. SNP cluster SalEnt-18-0030 is highlighted in blue as this contains the isolates associated with the outbreak commencing in May 2018.}
    \label{fig:time_SNP}
\end{figure}

\begin{figure}[h]
    \centering
    \includegraphics[width=\textwidth]{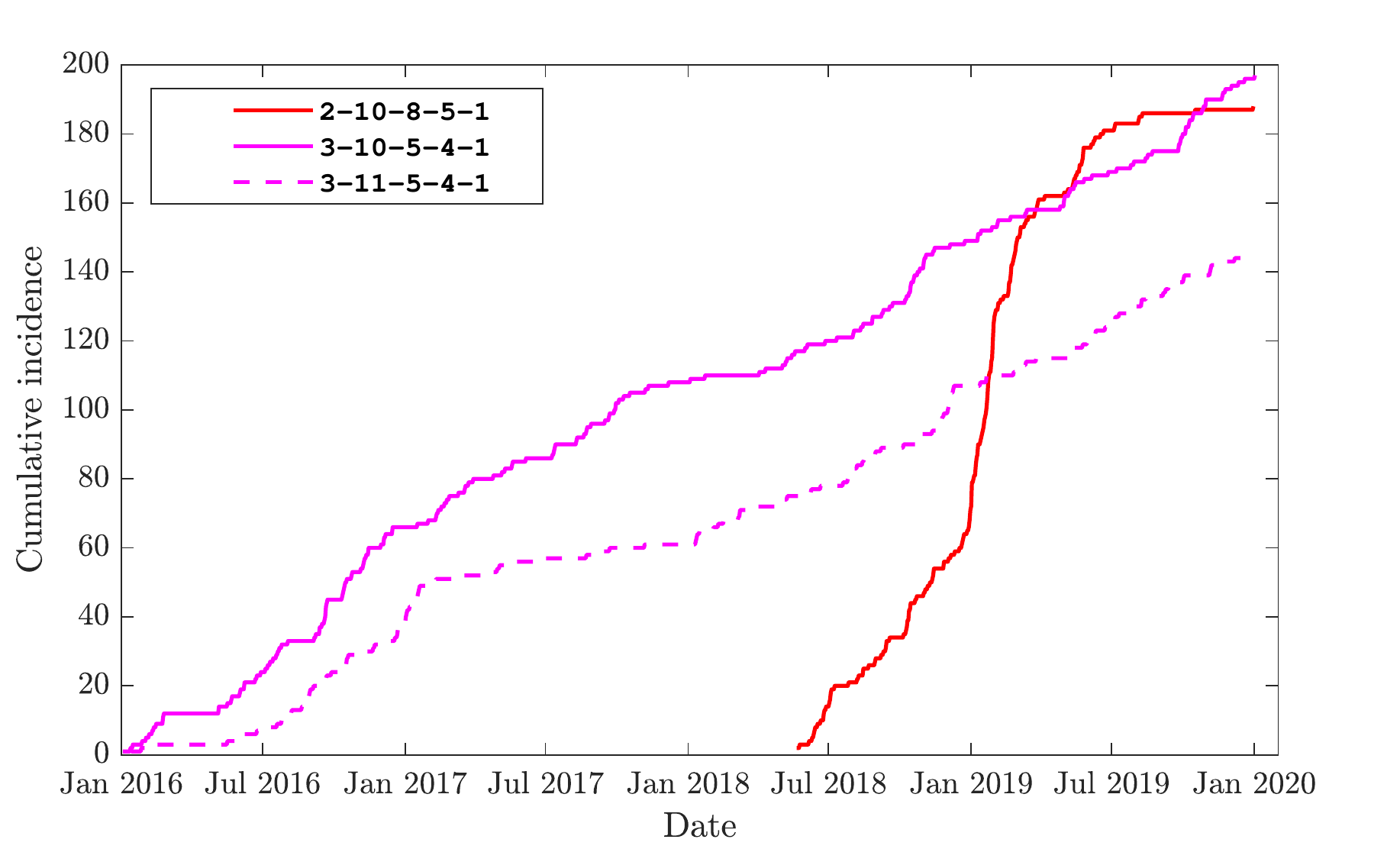}
    \caption{Cumulative incidence of the prevalent MLVA profiles highlighted in Fig.~\ref{fig:time_MLVA}.}
    \label{fig:inc_MLVA}
\end{figure}

There were 72 unique MLVA profiles and 78 unique SNP clusters (derived bioinformatically). The appearance of unique MLVA profiles and SNP clusters over time are shown in Figures \ref{fig:time_MLVA} and \ref{fig:time_SNP}. The majority of unique MLVAs and SNP clusters are sporadic. Notably, the MLVAs 3-10-5-4-1 and 3-11-5-4-1 appear consistently over the period of the data (both appear across SNP clusters SalEnt-16-0001 and SalEnt-16-0006). Of significance is the appearance of a strain responsible for a declared outbreak starting in May 2018 \parencite{nsw_health_salmonella_2018} (2-10-8-5-1, SalEnt-18-0030).            
Isolating these MLVAs on an epidemiological curve, illustrated in Figure \ref{fig:inc_MLVA}, shows the continual incidence of the persistent MLVAs 3-10-5-4-1 and 3-11-5-4-1, and the rapid increase in incidence for the outbreak associated 2-10-8-5-1.

\subsection{MLVA based network and centrality/prevalence space}

\subsubsection{MLVA network}
\label{mlva_net}
As a first step, we constructed a complete undirected network where each node is a unique MLVA and the edge weights are determined by Manhattan distance between the MLVAs ($N=72$, and the number of edges $M = N(N-1)/2 = 2,556$). Figure \ref{fig:undirected_net_MLVA} shows the thresholded undirected MLVA sub-network containing only those edges ($M=67$) with a distance of 1 (i.e., $G_{max}=1$). The persistent MLVAs 3-10-5-4-1 and 3-11-5-4-1 belong to a large connected component (shown in magenta, $n=31$). The outbreak forms a separate component around 2-10-8-5-1 (shown in black, $n=7$). We will refer to each component by the most prevalent MLVA, that is, 3-10-5-4-1 for the magenta component, 2-10-7-3-2 for the blue component and 2-10-8-5-1 for the black, outbreak-associated component. The 3-10-5-4-1 component and the outbreak 2-10-8-5-1 component represent 71\% and 22\% of total isolates respectively (636 and 192 isolates). The 2-10-7-3-2 component represents 3\% of total isolates (30 isolates), with the remaining 4\% of isolates (39 isolates) being disconnected singletons.

\subsubsection{MLVA centrality-prevalence space}

Following~\parencite{cliff_network_2019,cliff_inferring_2020}, the networks were used to relate the severity of an outbreak to the extent of niche exploitation by construction of the corresponding centrality-prevalence space. Closeness centrality as used in the centrality-prevalence space quantifies genetic connectivity across strains, distinguishing between peripheral and central nodes (e.g., genetic variation developing from central nodes can proliferate to the rest of the network). Thus, each node (either an MLVA profile or an isolate) was characterised in terms of two quantities: (a) the centrality of the node within the corresponding undirected genetic network, and  (b) its (average) genetic neighbourhood prevalence. In our previous construction of undirected MLVA network for the endemic \textit{S}. Typhimurium (STM) population in NSW over the period 2008-2016 \parencite{cliff_network_2019,cliff_inferring_2020}, the corresponding centrality-prevalence space revealed a structure that described salient evolutionary pathways. This included a region clearly delineating the dominant STM strains, as well as a transition region from which the dominant strains were most likely to originate. Thus, investigating this space allowed us to detect the role of each strain with respect to its emergent risk profile and the niche exploitation by the pathogen population. As argued in \parencite{cliff_network_2019,cliff_inferring_2020}, the region with moderate centrality but high prevalence is indicative of a niche exploitation by dominant, high-risk strains.

An MLVA based centrality-prevalence space was constructed for the 72 unique MLVAs of the \textit{S}. Enteritidis presentations in NSW. Figure \ref{fig:cen_MLVA} shows the centrality-prevalence space with a neighbourhood threshold distance of 1. The persistent MLVAs of 3-10-5-4-1 and 3-11-5-4-1 demonstrate both high centrality and prevalence. The outbreak-associated MLVA 2-10-8-5-1 shows both moderate centrality and prevalence. However, no clear structure emerges in this space, and consequently, the outbreak-associated MLVA 2-10-8-5-1 is not  clearly delineated in the MLVA based centrality-prevalence space.

\subsection{SNP network and centrality-prevalence space reveal outbreak strain} 

\subsubsection{SNP network}

Using SNP analysis on whole genome data improves the resolution for comparison down to the single base pair level. We constructed a complete undirected network where each node represents a single isolate and the edge weights are determined by SNP distance ($N=897$ and $M=401,856$). Figure \ref{fig:net_isolate_ud} shows the thresholded undirected SNP sub-network with only those edges with a distance of $\leq 20$ ($M=22,742$). In Figure \ref{fig:net_isolate_ud}, individual isolates are assigned a colour if the isolate was represented by one of the three MLVA components from Figure \ref{fig:undirected_net_MLVA}. Notably, the isolates in the 2-10-8-5-1 outbreak-associated component on the MLVA network form a single connected component on the SNP sub-network (shown in black, 196 isolates with 95\% concordance between undirected MLVA and SNP components). In contrast, the persistent component 3-10-5-4-1 of the MLVA network (shown in magenta) shows very little overlap between the networks, and instead forms 42 separate components of 3 isolates or greater and many pairs or singletons on the  SNP network (76\% of isolates in a component of $\geq 3$; the largest component with 66 isolates), indicating significantly greater heterogeneity at the  SNP level versus MLVA for these isolates. Similarly, the 2-10-7-3-2 component (shown in blue) of the MLVA network forms 3 components of 3 isolates or greater and 12 pairs or singletons on the SNP network (47\% in a component of $\geq 3$). 

In addition, we constructed a directed SNP sub-network including only the edges between nodes (isolates) separated by a SNP distance shorter than the threshold $G_{max} = 20$ and occurring within $T_{max} = 30$ days of each other ($M=4,252$). Figure \ref{fig:net_isolate_d} visualises this sub-network, again colour coding three MLVA components shown in Figure \ref{fig:undirected_net_MLVA}. Similarly to the undirected SNP sub-network, most of the isolates in the 2-10-8-5-1 outbreak-associated component of the MLVA network  form the largest connected component in the directed SNP sub-network as well (shown in black, 191 isolates with 92\% concordance between undirected MLVA and directed SNP components; maximal directed path length is 17; average directed path length is $4\mathord{\cdot}54$)\footnote{Strictly speaking, a connected component in any directed network is called a ``weakly connected component'', with its nodes connected to each other by some path, disregarding the direction of edges.}. Importantly, the persistent component 3-10-5-4-1 of the MLVA network (shown in magenta) fragments even further on the directed SNP sub-network, forming 41 separate components of 3 isolates or greater, with the largest magenta component containing only 12 isolates (maximal directed path length is 8; average directed path length is $3\mathord{\cdot}44$). This fragmentation occurs due to the significant temporal spread of these isolates in addition to their genetic heterogeneity. Furthermore, the 2-10-7-3-2 component (shown in blue) of the MLVA network mostly disintegrates into singletons, with one triplet and one pair.

\subsubsection{SNP centrality-prevalence space}

We constructed a centrality-prevalence space for the undirected network based on SNP distance. As above, we were seeking to observe a structure in this space delineating the isolates associated with the outbreak. Figure \ref{fig:cen_iso} shows this space, with each point representing a single isolate, and the isolates in the outbreak-associated network component (black) are well-delineated within a distinct region at high prevalence and moderate centrality relative to other isolates. This is indicative of a high-risk pathogenic population,  as observed in our previous work which identified dominant STM strains exploiting a niche~\parencite{cliff_network_2019, cliff_inferring_2020}. In contrast, the isolates that form a component in the thresholded undirected sub-network with edge threshold $\leq 20$ and not associated with the outbreak (blue) predominantly occupy a region of high centrality and moderate-low prevalence, indicative of not yet fully developed niche exploitation.

\subsection{Genetic distance network components compared to SNP cluster and phylogenetic analysis}

We sought to place the components which were derived within the MLVA and SNP based thresholded networks in context with the current standard methodologies for typing Salmonella populations. With a focus on the outbreak associated isolates, we compared four different typing methods: (i) MLVA alone, (ii) Components of the thresholded MLVA sub-network, (iii) Components of the thresholded SNP sub-network, (iv) SNP clusters derived from phylogenetic analysis. Figure \ref{fig:epi_com} traces an epidemiological curve showing the outbreak-associated 2-10-8-5-1 MLVA alone (shown in red, 188 isolates), the 2-10-8-5-1 MLVA network component (black, 194 isolates), the SNP cluster containing 2-10-8-5-1 isolates (green, 196 isolates) and the outbreak component in the SNP network (yellow, 196 isolates). In Table \ref{table:1}, we show the concordance between the SNP cluster SalEnt-18-0030 and the other typing methods.

With regards to the concordance of MLVA profiles 2-10-8-5-1 and SNP cluster SalEnt-18-0030, the correlation of MLVA profiles and SNP clusters was further investigated by compiling the correlations of the individual MLVA loci and all combinations of loci to the SNP distance (see Supplementary results and Supplementary figure \ref{fig:SNP_MLVA_corr}). 

To understand the components of the  SNP sub-network within the context of the population structure, we placed the components alongside a phylogenetic tree representing the \textit{S}. Enteritidis population as observed in this data set. As shown in Figure  \ref{fig:phylo}, the phylogenetic tree represents individual isolates and highlights the most prevalent SNP clusters and network component form a unique clade. Grey colouring in the metabar highlights cases that are part of the smaller SNP clusters ($k=100$), and isolates that do not cluster with other cases. The right meta bar highlights components of the network with greater than or equal to 10 members. The outbreak cluster has been enlarged (Figure \ref{fig:phylo}B) and concordances are visualised in the metabar between the SNP cluster, (191 isolates highlighted in blue), network component (193 isolates highlighted in black) and MLVA profile (187 isolates highlighted in orange).

\section{Discussion and conclusions}
\label{discussion}

Effective monitoring and investigation of emerging high-risk strains of \textit{S}. Enteritidis is of significant value to public health. In this study we examined data for \textit{S}. Enteritidis isolates causing foodborne Salmonella enterocolitis in NSW over the period of 2015-2019, including a major outbreak commencing in May 2018. In NSW, \textit{S}. Enteritidis is not an endemic pathogen and as such the data was expected to be highly heterogeneous. Specifically, it likely represented a sporadic sampling of diverse background populations from which the infections originated (i.e, importation of \textit{S}. Enteritidis into Australia). For each isolate, whole genome sequencing (WGS) was carried out. To examine the emergence of epidemic strains we constructed networks that were based on the relationships between (i) unique MLVA profiles and (ii) core genome SNPs of individual isolates. With these networks we constructed centrality-prevalence spaces that describe the relationships between unique MLVAs, as well as the individual isolates at the WGS level.

\begin{figure}[h]
    \centering
    \includegraphics[width=\textwidth,trim={2cm 1cm 2cm 1cm},clip]{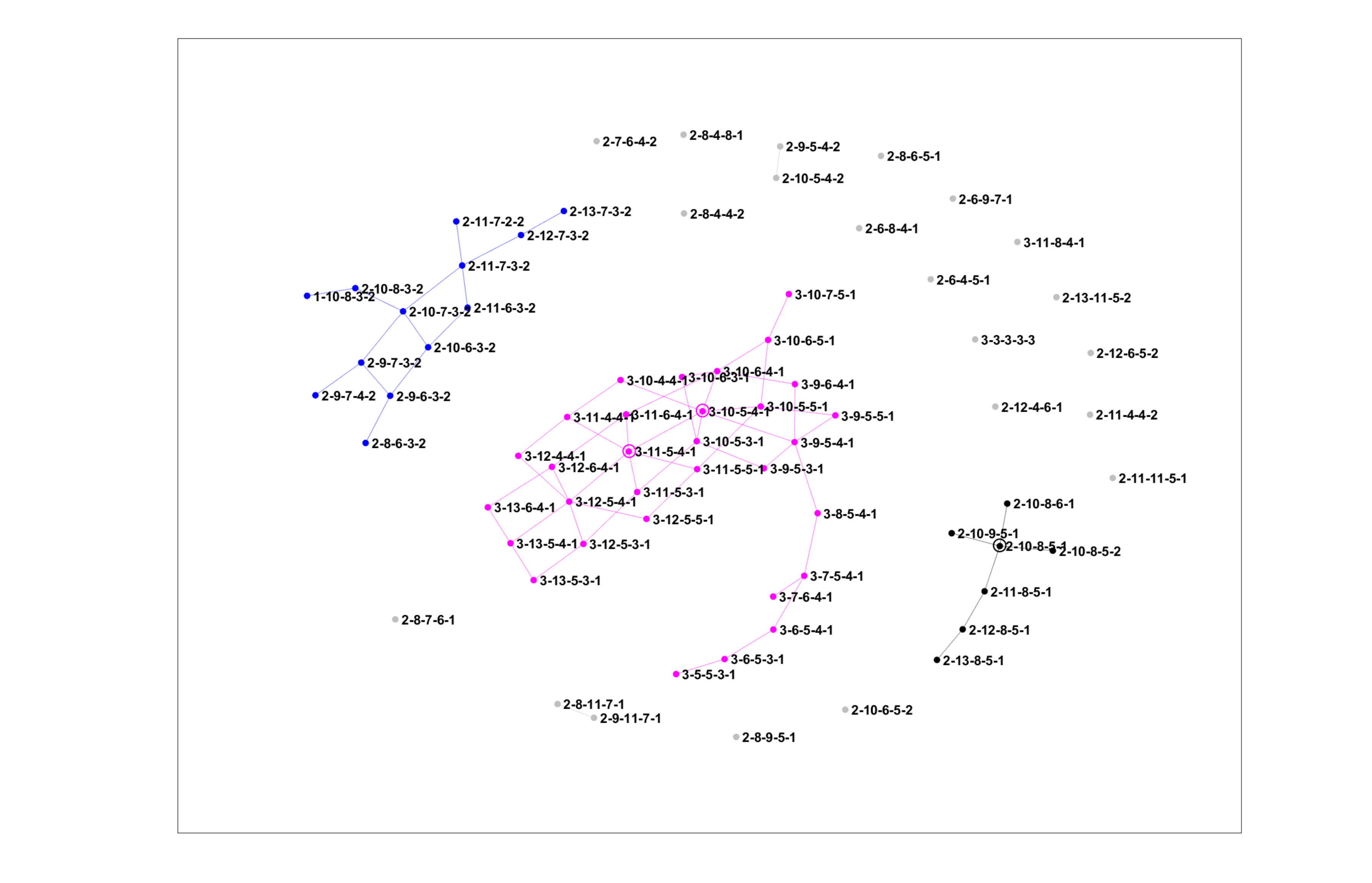}
    \caption{The undirected sub-network based on genetic distances between MLVA profiles. $N = 72$; $M = 67$. An edge is inferred if the genetic distance is 1 (i.e., $G_{max}= 1$). The nodes highlighted with a ring represent the persistent MLVA profiles 3-10-5-4-1 and 3-11-5-4-1 (magenta), as well as the outbreak MLVA profile  2-10-8-5-1 (black).}
    \label{fig:undirected_net_MLVA}
\end{figure}

\begin{figure}[h]
    \centering
    \includegraphics[width=\textwidth,trim={1cm 0cm 1cm 0cm},clip]{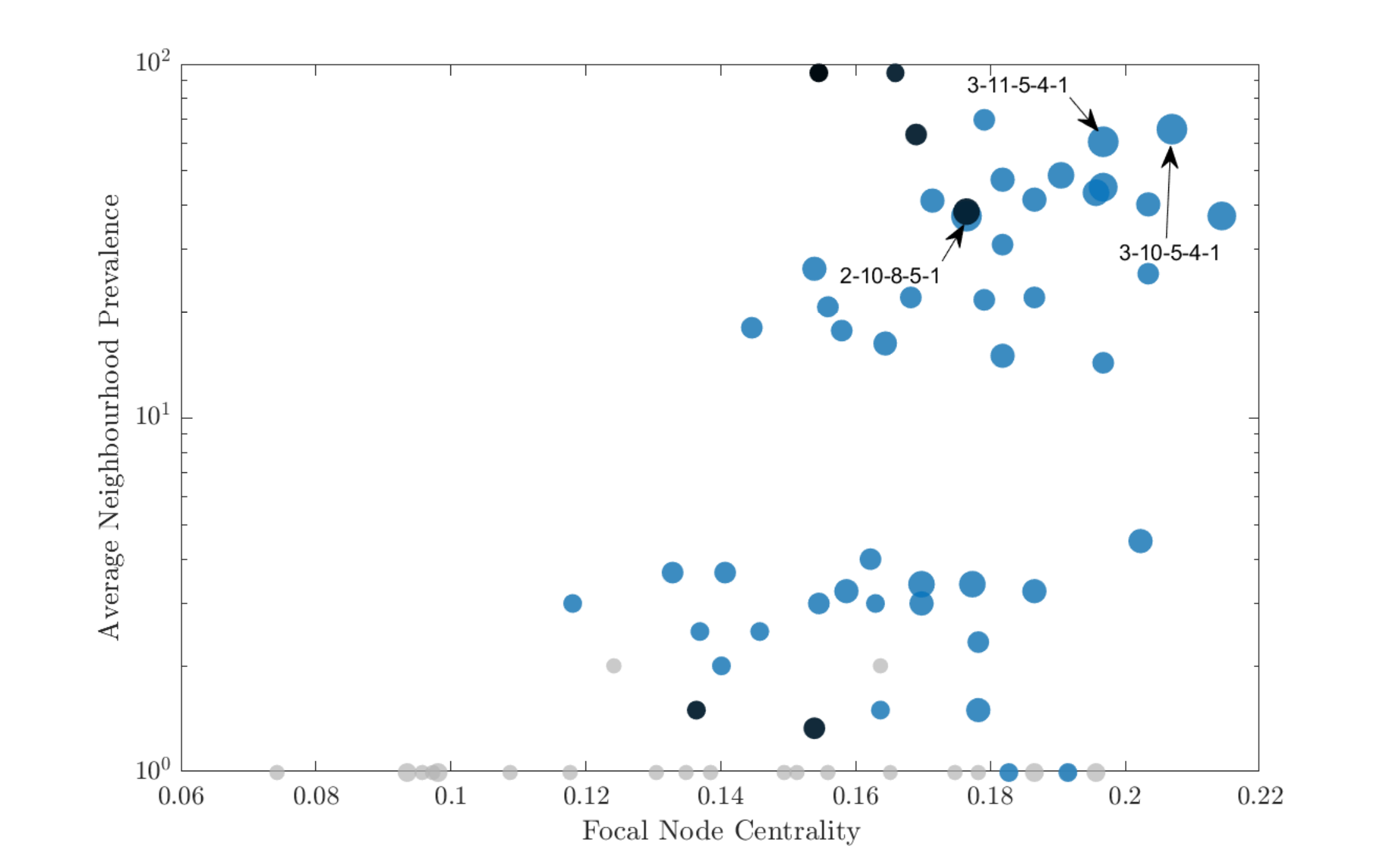}
    \caption{Centrality-prevalence space for the undirected MLVA network where each node represents a unique MLVA profile, mapping the centrality of profiles (x-axis) and their average neighbourhood prevalence (y-axis). Neighbourhood threshold $G_{max}=1$. The size of each point is proportional to the size of the corresponding neighbourhood (i.e., the number of profiles). Example: the overlapping neighbourhood of MLVA profile 2-10-8-5-1 contains four other MLVA profiles within genetic distance $G_{max}=1$ from it: 2-10-8-5-2, 2-10-8-6-1, 2-10-9-5-1 and 2-11-8-5-1, each of which is a point in the centrality-prevalence space.  Colour of the profiles corresponds to the colour of the network components shown in Fig.~\ref{fig:undirected_net_MLVA}, e.g., the nodes shown in blue belong to non-outbreak components (e.g., persistent MLVA profiles 3-10-5-4-1 and 3-11-5-4-1), and the nodes shown in black belong to the outbreak component (e.g., the outbreak MLVA profile 2-10-8-5-1.}
    \label{fig:cen_MLVA}
\end{figure}

\begin{figure}[h]
    \centering
    \includegraphics[width= \textwidth,trim={2cm 1.5cm 2cm 0.5cm},clip]{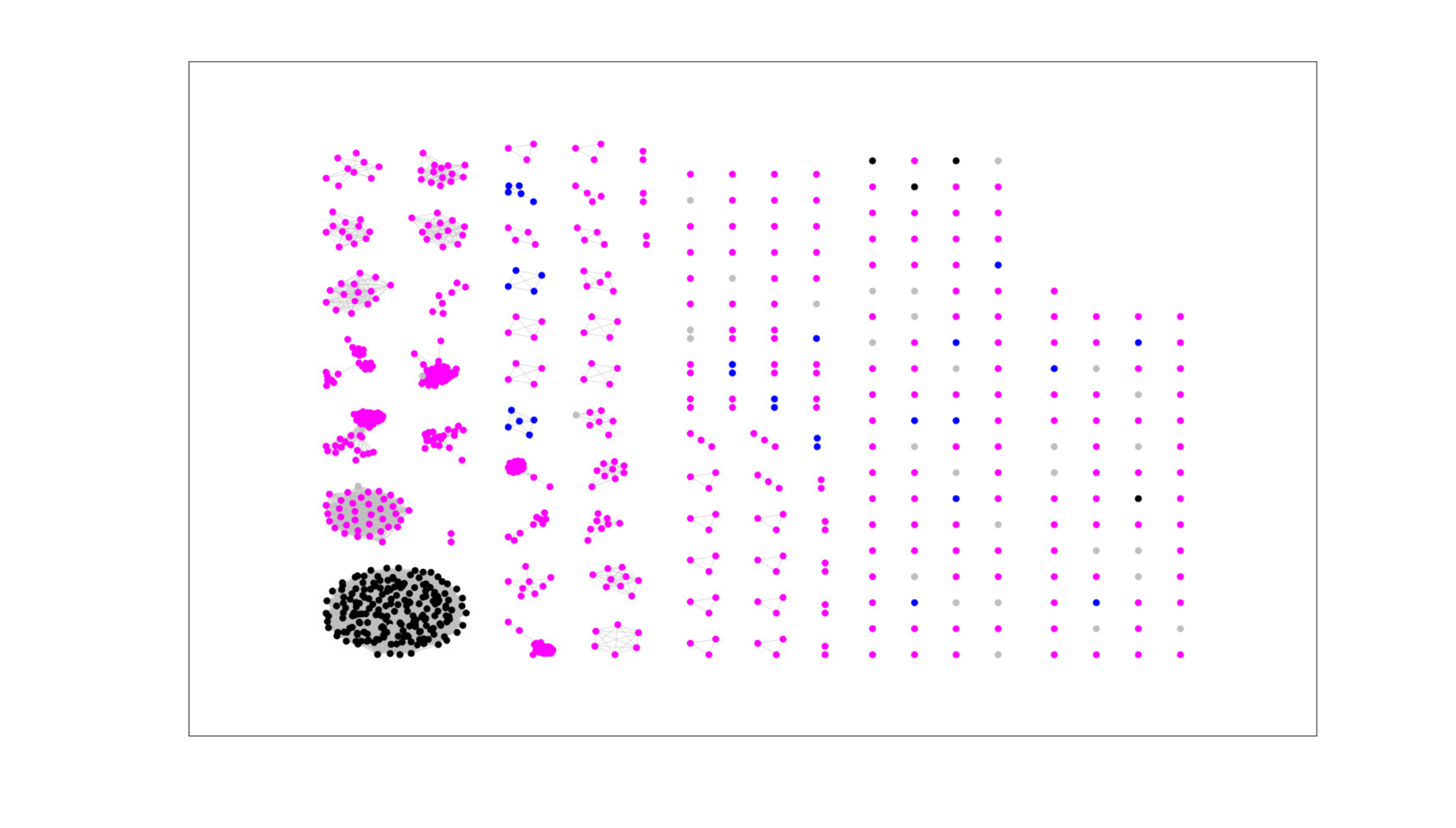}
    \caption{The undirected thresholded SNP sub-network (SNP distance threshold $G_{max}=20$;  $N=897$; $M=22,742$). Singletons: 162; total number of components: 229 (including singletons). Components highlighted in black, magenta, and blue corresponds to MLVA components identified in Fig.~\ref{fig:undirected_net_MLVA}.}
    \label{fig:net_isolate_ud}
\end{figure}

\begin{figure}[h]
    \centering
    \includegraphics[width= \textwidth,trim={2cm 1.5cm 2cm 0.5cm},clip]{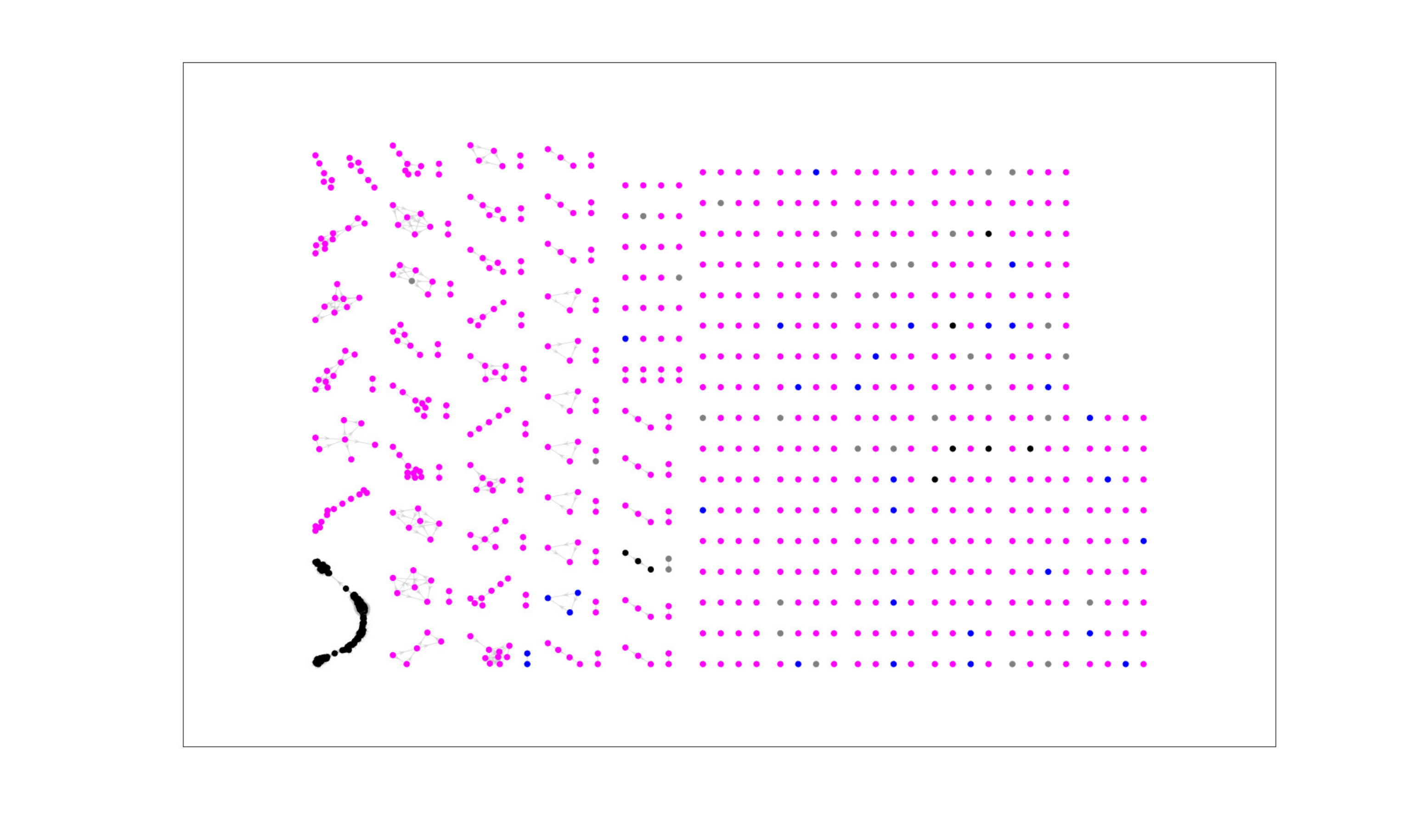}
    \caption{The directed SNP sub-network  (SNP distance threshold $G_{max}=20$; time window $T_{max} = 30$ days; $N=897$; $M=4,252$). Singletons: 400; total number of components: 483 (including singletons). Components highlighted in black, magenta, and blue corresponds to MLVA components identified in Fig.~\ref{fig:undirected_net_MLVA}.}
    \label{fig:net_isolate_d}
\end{figure}

\begin{figure}[h]
    \centering
    \includegraphics[width=\textwidth,trim={1cm 0cm 1cm 0cm},clip]{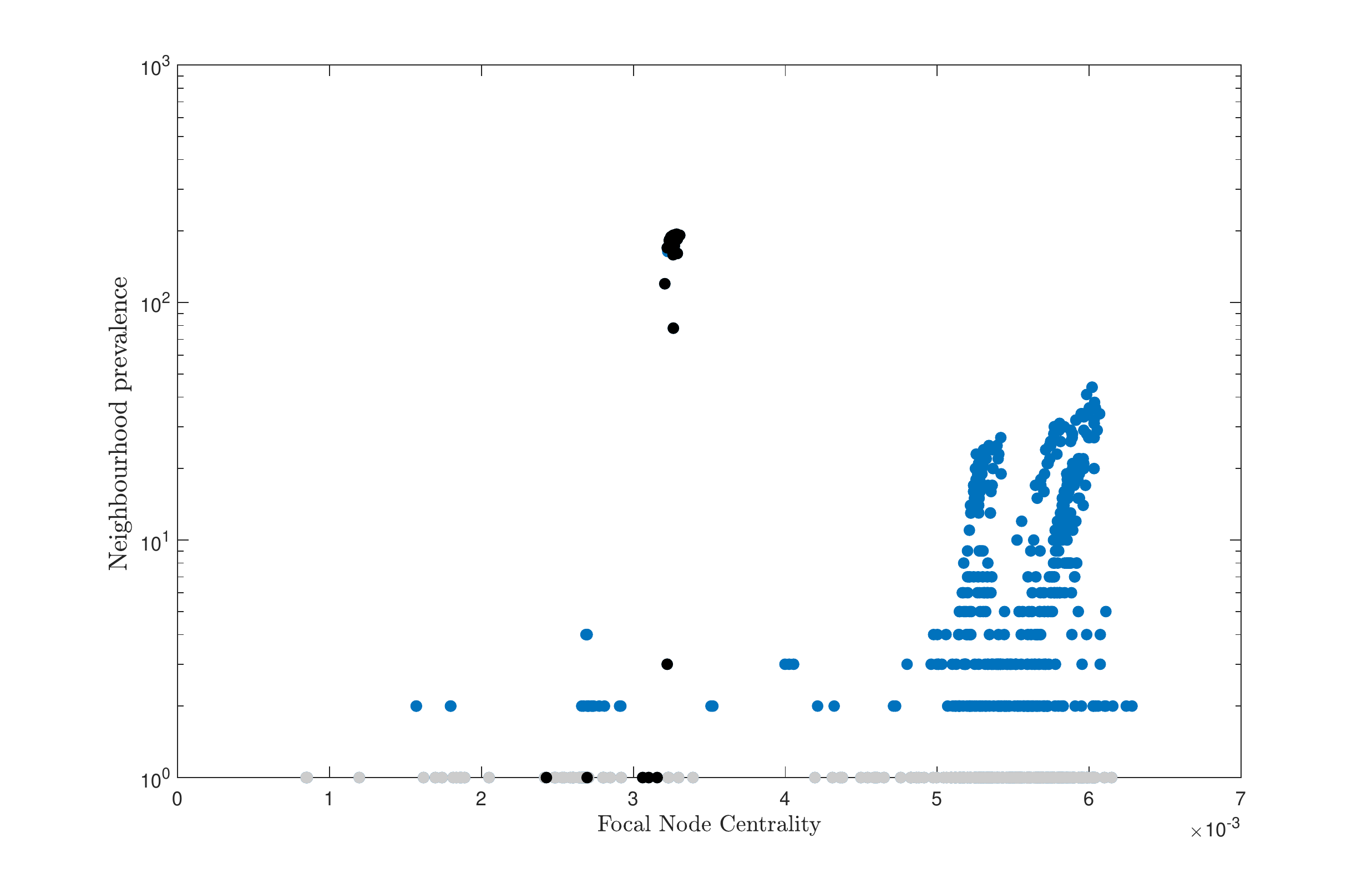}
    \caption{Centrality-prevalence space for the undirected SNP network where each node represents an isolate, mapping the centrality of profiles (x-axis) and their neighbourhood prevalence (y-axis). Neighbourhood threshold $G_{max}=20$. Disconnected nodes are shown in gray colour. Isolates highlighted in black belong to the component shown in black in Fig.~\ref{fig:undirected_net_MLVA}.  The nodes shown in blue belong to non-outbreak components. Note that since each isolate is unique, the neighbourhood prevalence here is interpreted as the neighbourhood size.}
    \label{fig:cen_iso}
\end{figure}

\begin{figure}[h]
    \centering
    \includegraphics[width= \textwidth]{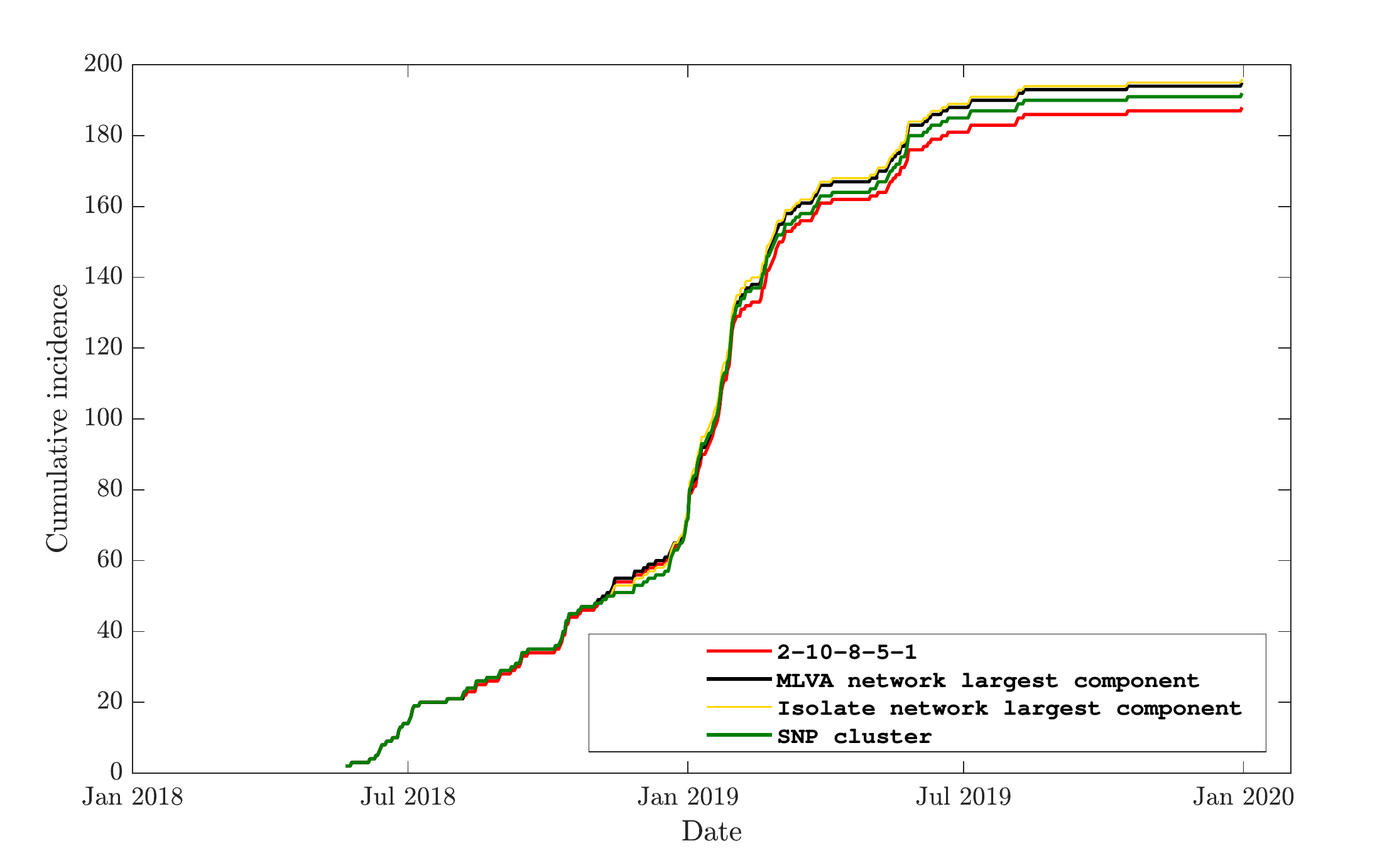}
    \caption{Cumulative incidence of the prevalent profiles identified in undirected MLVA network (black, Fig.~\ref{fig:undirected_net_MLVA}), SNP cluster SalEnt-18-0030 (green, Fig.~\ref{fig:time_SNP}), and undirected isolate network (yellow, Fig.~\ref{fig:net_isolate_ud}).}
    \label{fig:epi_com}
\end{figure}

\begin{figure}[h]
    \centering
    \includegraphics[scale=1.5] {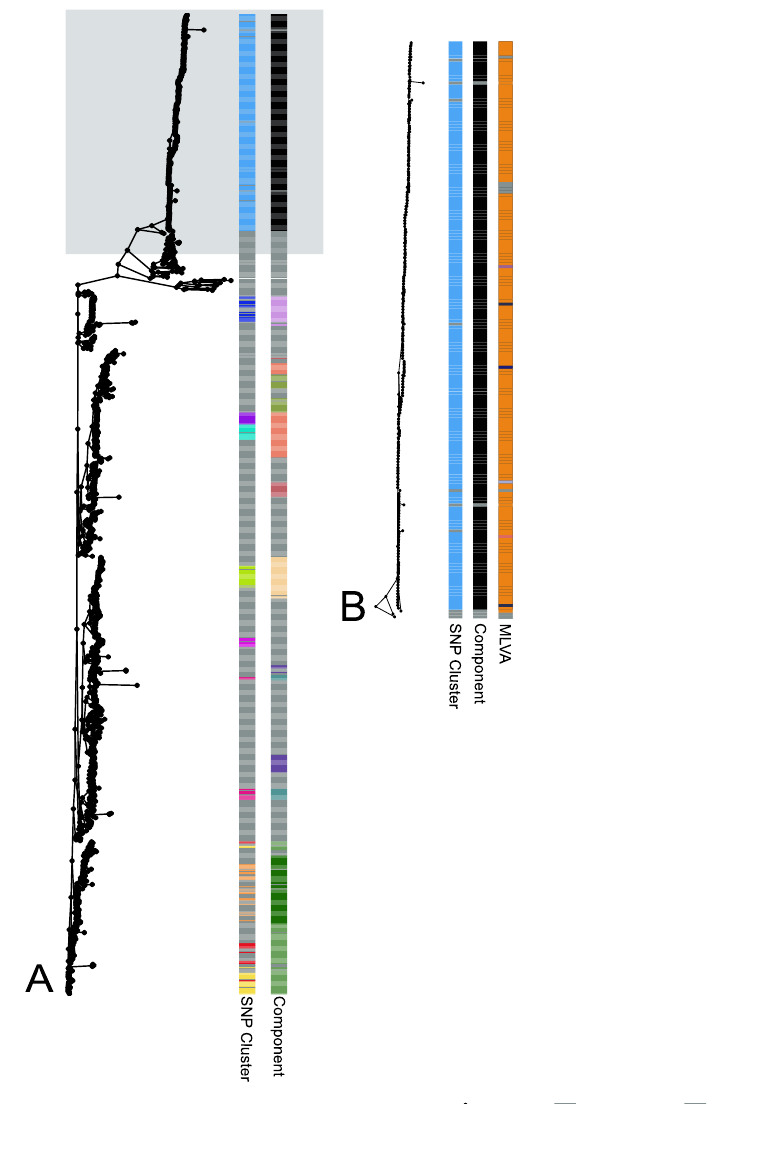}
    \caption{A. Maximum likelihood phylogeny built from the core genome of 897 Salmonella Enteritidis isolates. For each isolate, the metabars indicate if the isolate was a member of a SNP cluster and/or an undirected network component. Highlight colour on the left metabar indicates the 10 most prevalent SNP clusters, and the highlight colour on the right metabar indicates the 10 most prevalent undirected network components. The components and SNP cluster details associated with each colour are detailed in Supplemental Table \ref{fig:colour}.  All components and SNP clusters outside the 10 most prevalent respectively, as well as singletons, are indicated with grey on the metabars. B. A sub-tree of the outbreak clade, drawn from the region highlighted in the grey box of part A, demonstrating high concordance between undirected network components, SNP based clustering and MLVA for the outbreak. For each isolate, the highlight colour on the corresponding metabar indicates if the isolate was a member of a SNP cluster, undirected network component and/or MLVA profile from left to right respectively. An isolate which is not associated with one of these groupings is indicated by grey in the corresponding metabar. The details for components, SNP clusters and MLVA associated with each colour are detailed in Supplemental Table \ref{fig:colour}.} 
    \label{fig:phylo}
\end{figure}

The network structure demonstrated a higher resolution achievable by SNP distance between individual isolates than by MLVA. The undirected MLVA network showed three distinct components. The largest component (Figure \ref{fig:undirected_net_MLVA}, magenta) contained the `persistent' MLVAs, such as 3-10-5-4-1, which were observed repeatedly over the time course of the data set and represent the bulk (71\%) of isolates. The membership of these MLVAs within a single large component, formed at a minimal genetic distance $G_{max}=1$ on the thresholded MLVA sub-network, indicates their close genetic proximity, with small diversification. If considered in isolation, this may have been suggestive of an existing endemic population of strains which lead to locally acquired infections, so that a sufficiently frequent sampling has captured local diversification. However, the higher resolution of the SNP sub-networks invalidates this hypothesis. Specifically, in the thresholded undirected SNP sub-network, the 3-10-5-4-1 MLVA component is broken up into a multitude of small connected components, including just under a quarter of these previously connected isolates as pairs or singletons. Moreover, the directed SNP sub-network fragments the connectivity of these isolates even further, highlighting their spread over time in addition to their genetic diversity. That is, the isolates genetically close to 3-10-5-4-1 MLVA tend to form short directed paths on SNP sub-network, indicating that the genetic changes usually are not fixed in subsequent generations in \textit{Salmonella} Enteritidis. This reveals that there were significant differences between these isolates at the SNP level that were not captured by MLVA. Thus, the improved resolution afforded by SNP distance and captured by the SNP networks suggests sporadic sampling from a diversity of infection sources, rather than endemic population. In other words, this points to multiple, and often concurrent, importation routes for this set of \textit{S}. Enteritidis infections (i.e., proportions of infections were acquired overseas but diagnosed and recorded in Australia as well as could be explained by exposure to diverse food sources contaminated with the pathogen).

On the other hand, the outbreak-associated component, formed on the MLVA network around 2-10-8-5-1 (Figure \ref{fig:undirected_net_MLVA}, black), may have been suggestive of locally acquired infections. In this case, the thresholded SNP sub-network supported this conclusion, with the majority (98\%) of these isolates grouped into a single connected component.  That is, there is a strong genetic proximity  --- at the SNP level --- within this large component, ruling out multiple concurrent importation routes. The connectivity is reinforced by the directed SNP sub-network,  grouping these isolates (96\%) in the largest directed component which signifies their genetic as well as temporal proximity. This strongly suggests frequent sampling of a single imported  strain which subsequently resulted in a large number of locally acquired infections.

The centrality-prevalence space constructed for SNP network clearly delineated the outbreak isolates. In our previous work we have described a transition region of moderate prevalence and high centrality, from which outbreak strains can be observed to transition to high prevalence and moderate centrality \parencite{cliff_network_2019, cliff_inferring_2020}. The centrality-prevalence space based on the MLVA network does position the 2-10-8-5-1 component at a moderate centrality, but the relative diversity of MLVAs with single or low prevalence in this component resulted in a moderate centrality/moderate prevalence position. In contrast, the improved resolution of the SNP network resulted in a strong delineation of the outbreak isolates into a `high-risk signature' with a moderate centrality/high prevalence position. The moderate centrality is a result of each isolate in this region being moderately genetically close to the rest of the network while at the same time forming an emerging sub-population. As such, it is the intensity of the outbreak and its acquired genetic prominence relative to the background population which shapes its separation in the centrality/prevalence space. An important observation from these findings is that, with sporadic data such as that for imported \textit{S}. Enteritidis, identifying a high-risk strain can be achieved much earlier in an outbreak based on the centrality of the first identified component containing the outbreak isolates.

In this study we examined the population clustering of food-borne bacteria by the gold standard approaches of SNP-based clustering using phylogenetic analysis in comparison to clusters determined as components on either the MLVA or pairwise SNP count based networks. A comprehensive comparison between phylogenetic trees and genotype networks is outside of scope for our study, with the relative merits of these approaches being dependent on context \parencite{bapteste_towards_2018, blais_past_2021}. 
With respect to the outbreak, the network components and SNP clustering all performed similarly, identifying the main bulk of the isolates under MLVA 2-10-8-5-1 and the related isolates within the outbreak. This correlation is notable, as the relative complexity of the network approach is much less than that of phylogenetic inference and analysis. In addition, a phylogenetic tree needs to be routinely updated and adjusted, particularly when a novel or previously unidentified strain emerges.

In the context of sporadic and heterogeneous data typical of epidemiological surveillance, our study provides strong evidence that network analysis of WGS sequencing data is complementary to traditional phylogenetic analysis. We believe that network analysis as a companion tool to phylogenetic analysis improves the resolution of population and outbreak analysis. When considered in terms of complexity, the network approach is superior to traditional phylogenetic analysis in defining outbreaks that require substantive and rapid public health investigation and action. As has been noted in particular for analysing bacterial populations, a phylogenetic tree will always produce a hierarchically connected structure which may not accurately represent a bacterial population \parencite{bapteste_prokaryotic_2009}, especially a population observed during disease surveillance. A network based on sporadic, heterogeneous genomic data will produce an appropriately disconnected network structure. In addition, a network may be the better representation of the lateral movement of genetic material \parencite{corel_network-thinking_2016, bapteste_towards_2018}, a phenomenon noted with respect to virulence factors \parencite{nieto_new_2016, dos_santos_virulence_2019}, that make hierarchical representation of bacterial populations difficult. While the network approach here successfully delineated the outbreak isolates, further investigation into the accuracy of network components as a more general tool for population typing may be of utility in the context of \textit{Salmonella} spp. surveillance. 

Some limitations of this study have to be acknowledged. Our network analyses, and that of the gold-standard phylogenetic analysis, are based on the core genome and as such do not take into account the information contained in the larger accessory genome space. Effective use of this much more fluid genetic information would enable a significantly higher resolution for accurate partitioning of the population, as well as enable investigation of genetic variations that bring about high-risk strains and outbreaks. Utilising the information of the accessory genome is the next step in network based genomic analysis of pathogenic bacterial populations.

In conclusion, network based analysis of a genome level data derived from bacterial isolates can improve the resolution of population analysis and public health surveillance. Specifically, the outbreak of \textit{S.} Enteritidis was confidently delineated from the rest of the bacterial population on both the undirected SNP based network and the associated centrality/prevalence space. The improved resolution at the SNP level and incorporation of time of acquisition revealed the sub-components within the frequently observed ‘persistent’ SNP cluster and MLVA patterns, which gives greater clarity to the determination of the origin of \textit{S.} Enteritidis infections. High concordance of network components and SNP clusters is a promising insight for development of rapid population analyses of foodborne \textit{Salmonella} spp. population due to the low overhead of network analysis.

\subsection*{Declaration of interest}
The authors declare that they have no competing interests.

\subsection*{Acknowledgements}

NSW Health Pathology Enteric Reference Laboratory and NSW Health Pathology Public Health Pathogen Genomics Laboratory

\subsection*{Contributions}
AS, SC, RR, OC, TS, VS and MP designed the study and analysed the data. RR, QW, AA and MR collected and curated the data. AS, SC, RR, TS, VS and MP wrote and edited the manuscript. AS, SC, RR, QW and AA had access to and verified the underlying data and approved the final version of the manuscript. All authors reviewed the final draft and agree with its content and conclusions.

\subsection*{Funding}
This work was supported by the Australian Research Council grant DP200103005 (AS, SC, TS, VS and MP). 
 
\newpage

\printbibliography

\newpage
\pagenumbering{gobble}

\begin{table}[h]
    \centering
    \begin{tabular}{| c || c |}
        \hline
        \textbf{SNP cluster SalEnt-18-0030 vs.} & \textbf{Concordance} \\  \hline
        \textbf{Set of isolates with MLVA 2-10-8-5-1} & $0.91$ \\ \textbf{Largest component of MLVA undirected sub-network } & $0.93$ \\
        \textbf{Largest component of  undirected SNP sub-network} & $0.98$ \\
        \textbf{Largest component of  directed SNP sub-network} & $0.95$ \\
        \hline
    \end{tabular}
    \caption{Concordance between the outbreak-associated SNP cluster SalEnt-18-0030 and the outbreak-associated MLVA (2-10-8-5-1), as well as MLVA and SNP network components.}
    \label{table:1}
\end{table}

\newpage

\newpage

\newpage

\newpage

\newpage

\newpage

\newpage

\newpage

\newpage

\newpage

\clearpage

\beginsupplement
\section{Supplementary}

\subsection{Supplementary methods}
\subsubsection{Isolate collection}
The \textit{S. enterica} subspecies \textit{enterica} serovar Enteritidis clinical isolates included in this study represent all isolates referred to the NSW Enteric Reference Laboratory, Institute of Clinical Pathology and Medical Research (ICPMR), NSW Health Pathology. All referred isolates underwent whole-genome sequencing as part of routine public health outbreak investigations in NSW, Australia between August 2015 and December 2019 (1033 isolates).

\subsubsection{Nucleic acid extraction and library preparation}
A single colony was used for DNA extraction; extraction was performed using the Geneaid Presto genomic DNA bacterial kit (Geneaid, Taiwan) as per the manufacturer’s instructions for Gram-negative bacteria. DNA extracts were treated with 1U of RNase. DNA libraries were prepared with the Nextera XT library preparation kit, using 1 ng of DNA in accordance with manufacturer’s instructions. Multiplexed libraries were sequenced using paired-end 150-bp chemistry on the NextSeq 500 system (Illumina, Australia).

\subsubsection{Bioinformatic analysis of sequenced genomes}
Demultiplexed sequencing reads with $ >1 \times 10^7$ reads per isolate were trimmed \parencite{bolger_trimmomatic_2014}, based on a minimum quality read score of 20, and then \textit{de novo} assembled using SPAdes (version 3.13.0) \parencite{bankevich_spades_2012}. The quality of \textit{de novo }assemblies was assessed with Quast (version 5.0.2) \parencite{gurevich_quast_2013}; only assemblies with $<200$ contigs and N50 values of $>50,000$ bp were included in further analysis.

Multiple-locus variable-number tandem-repeats analysis (MLVA) was carried out from final contigs using  MLVA In Silico Typing Resource for Salmonella Strains (MISTReSS, https://github.com/Papos92/MISTReSS). WGS has become the gold standard typing method to investigate foodborne disease outbreaks, however it is still not universally used internationally, therefore numerous studies have validated the backwards compatibility of in silico MLVA typing to enable international comparisons \parencite{ambroise_backward_2019, sacchini_whole_2019, holzer_tracking_2021}.    

Contigs were annotated with Prokka (version 1.13.3) \parencite{seemann_prokka_2014}. Core genome analysis was determined using Roary (the pangenome pipeline; version 3.12.0) with default parameters \parencite{page_roary_2015}. SNP differences between isolates were quantified from the core genome alignment using SNP-sites (https://github.com/sanger-pathogens/snp-sites) \parencite{page_snp-sites_2016} without recombination masking \parencite{gorrie_key_2021}. Maximum likelihood phylogeny of the core genome was generated with iqtree using a full alignment, model finder and 1000 ultra-fast bootstraps (version 1.6.7) \parencite{nguyen_iq-tree_2015}. Tree annotation was conducted with the R package ggtree (version 1.14.6) \parencite{yu_ggtree_2017}. Nine reference isolates were included in the analysis to represent the three major \textit{S.} Enteritidis lineages detected in Australia (Supplementary figure \ref{fig:Max_likelihood_phylo}).

The relationship between isolates was quantified by two methods: (i) an \textit{in silico} estimation for MLVA over 5 loci, and (ii) quantification of SNPs within the core genome defined as the portion of the pangenome present in $\geq 99 \%$ of isolates. SNP clusters were generated in order of appearance based on genomic distance (<10 SNPs) to an index isolate in each cluster. 

SNP thresholds in this study were based on SNP distances observed in historical WGS data from \textit{S.} Enteritidis outbreaks. \textcite{inns_prospective_2017} identified the median pairwise distance between \textit{S.} Enteritidis isolates of an outbreak to be 10 SNPs, with a max of 20 SNPs. \textcite{taylor_angela_j_characterization_2015} similarly observed an upper threshold of 20 SNPs between isolates of a \textit{S.} Enteritidis outbreak.

\subsubsection{Network construction and centrality-prevalence space analysis}

We constructed two empirical undirected networks (complete graphs): (i) MLVA network and (ii) SNP network. In the MLVA network each node represented a distinct MLVA profile, and the edge weight between MLVA nodes was defined as the Manhattan distance $G_{ij}$ between MLVA nodes $i$ and $j$. For the SNP network each node corresponded to a unique isolate, and the edge weights were determined by the pairwise SNP distances $G_{ij}$ between isolates $i$ and $j$. 

For both the MLVA and SNP based networks, we derived an overlapping neighbourhood for each node by including all neighbour nodes within a certain distance from the node~\parencite{cliff_network_2019,cliff_inferring_2020}. Specifically, the neighbourhood $C_i$ associated with node $i$ is given by $C_i = \{j: G_{ij} \le G_{max}\}$, for some distance threshold $G_{max}$. For the MLVA network $G_{max} = 1$ which provides the highest resolution. For the SNP network $G_{max} = 20$.

Our intention is to compare the current standard of SNP clustering with the group structure of the SNP-based network, by comparison of SNP clusters to network components. The difference in thresholds for (a) SNP clusters derived chronologically based on genomic distance (<10 SNPs) and (b) components on the SNP network ($G_{max} = 20$) is due to the nature of clustering. SNP clusters are constructed on the distance between each isolate and an index isolate, making the distance analogous to the radius of the cluster. In contrast, the SNP network components are constructed using the pairwise distance between isolates, making the distance analogous to the diameter of the component. Informally, the overlapping network neighbourhoods approximated genetic neighbourhoods of each strain --- either in terms of MLVA or SNP. We note that the overlapping (network) neighbourhoods inferred in the SNP network differ, in general, from the SNP clusters derived bioinformatically.  

The closeness centrality~\parencite{bavelas1950communication,sabidussi1966centrality} of the centrality-prevalence spaces was normalised across $N$ nodes, and computed for each node as follows: 
\[l_i =  \frac{N}{\sum_{j=1, i \neq j}^{N} G_{ij}} .\]
The quantity $l_i$ quantifies the average shortest distance from node $i$ to each other node, thus capturing the relative importance of the node $i$ in terms of its proximity to other nodes. The average of the  inverse normalized  closeness  centrality  $\lambda_i = 1/l_i$ is equal to the average  path length $L$ which characterises the extent of genetic variation in the population on average:
\[L =  \frac{1}{N(N-1)}{\sum_{i=1}^N \sum_{\substack{j=1 \\ i \neq j}}^N G_{ij}} = \frac{1}{N-1} \sum_{i=1}^N \left ( \frac{1}{N} \sum_{\substack{j=1 \\ i \neq j}}^N G_{ij} \right ) =  \frac{1}{N-1} \sum_{i=1}^N \frac{1}{l_i} = \frac {1} {N-1 } \sum_{i=1}^N \lambda_i .\] 
This relation to the average path length motivated our choice of the closeness centrality, rather than betweenness centrality~\parencite{freeman1977set} or eigenvector centrality~\parencite{zaki2014data}. This empirical choice simply emphasises our focus on the average length of the shortest paths, rather than their number, allowing us to measure the  contribution of each node to the average  genetic variation in the population.

In order to characterize the severity of outbreaks within the MLVA network,  we computed the average prevalence for each overlapping neighbourhood. This average was obtained by dividing the number of all occurrences of the MLVA profiles contained within the neighbourhood by the neighbourhood size (i.e., accounting for all isolates with these MLVA profiles). For the SNP network, the neighbourhood prevalence was obtained instead as the neighbourhood size (since each isolate is unique, and the average neighbourhood prevalence is equal to 1).

In addition to the complete undirected MLVA and SNP networks, we derived their thresholded sub-networks, retaining only the edges with distances not exceeding the corresponding  threshold $G_{max}$. We used weight thresholding as one of the simplest and robust approaches to graph sparsification, found to be preserving group structure in real-world networks~\parencite{yan2018weight}. The sensitivity analysis shows that the network group structure and the structure of the centrality-prevalence space are robust to changes in $G_{max}$, see Supplementary Figures \ref{fig:Fig6_sensi}, \ref{fig:Fig7_sensi}, \ref{fig:Fig8_sensi}, and Tables \ref{fig:undirected_network_properties} and \ref{fig:directed_network_properties}. The connected components of these sub-networks were contrasted with the overlapping MLVA neighbourhoods, as well as the bioinformatically defined SNP clusters.  A comparison between any two sets $X$ and $Y$ was carried out by computing a concordance measure which contrasted sizes of the sets' intersection and union: $\frac{|X \cap Y|}{|X \cup Y|}$, where $|\cdot|$ denote the set size. 

Finally, we constructed a directed SNP sub-network which captured both genetic and temporal proximity. Specifically, a directed edge between two nodes (i.e., isolates) is inferred when their SNP distance is within the threshold $G_{max}$ and their detection dates are within a fixed time window $T_{max}$, with the edge direction determined by the temporal precedence (if the dates are the same, two directional edges are inferred). A directed path, formed by connected directed edges, approximates a chain of successive (potential) adaptation steps~\parencite{cliff_inferring_2020}. For the directed SNP network, $T_{max}$ = 30 was determined as reflecting the median shedding period associated with foodborne salmonellosis \parencite{medus_salmonella_2006}. Additionally, salmonellosis is characterised by acute symptom onset within 6 hours - 6 days \parencite{medus_salmonella_2006}, which provides confidence that date of detection is a reliable approximation of the date of infection in this context.

\subsection{Correlation of MLVA and SNP metrics for \textit{S}. Enteritidis genetic distance}

\begin{figure}[h]
    \centering
    \includegraphics[scale=0.6]{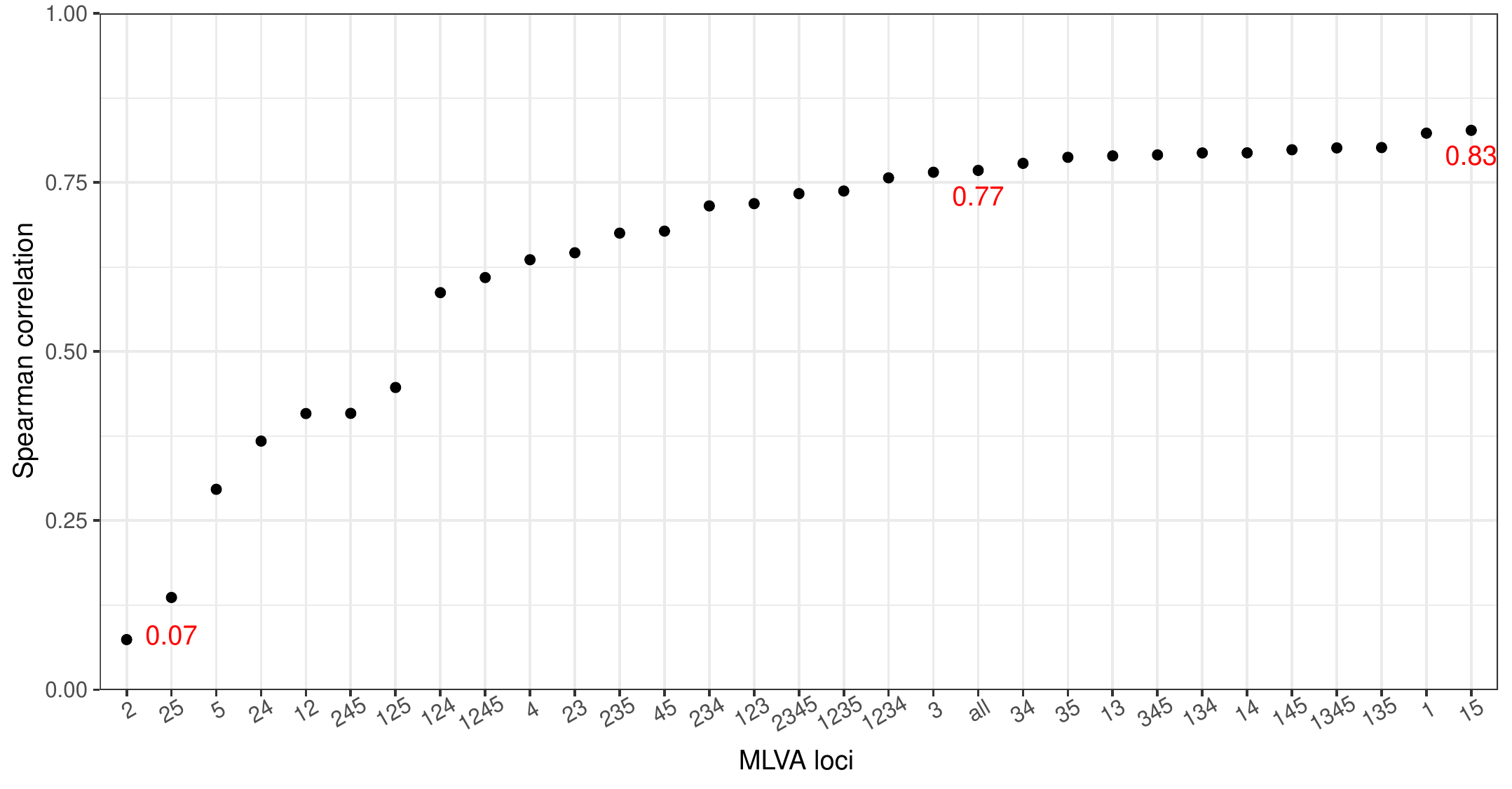}
    \caption{Spearman correlation of MLVA loci and core SNP distance.}
    \label{fig:SNP_MLVA_corr}
\end{figure}

\begin{figure}[h]
    \centering
    \includegraphics[scale=0.7]{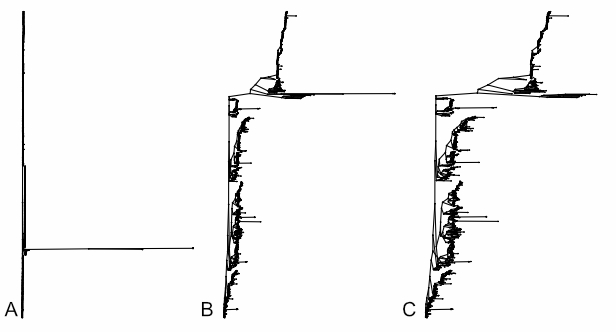}
    \caption{Maximum likelihood phylogeny of the core genome of Salmonella Enteritidis. A. Demonstrates the genomic diversity of the three major Salmonella Enteritidis lineages, nine reference genomes were included in the analysis. Lineages II and III are represented by divergent branches. B. Only represents isolates in Lineage I which predominates in NSW, four isolates still demonstrate significant genetic diversity and confound the resolution of closely related genomes. C. Divergent lineage I isolates are collapsed easing the interpretation of relatedness of genomes, this phylogeny is represented in the main text.}
    \label{fig:Max_likelihood_phylo}
\end{figure}

MLVA is a widespread typing tool for \textit{Salmonella} spp. Each MLVA profile comprises the variable number of tandem repeats across five loci. Typically, MLVA profiles differ between bacterial strains but remain similar for epidemiologically linked cases. As a result, they offer some discriminatory power during outbreak investigations. SNP counts within the core genome for every isolate pair (401,856 pairwise scores) were calculated and the Spearman correlation between pairwise MLVA distance and SNP distance across all isolates was determined (Supplementary Figure  \ref{fig:SNP_MLVA_corr}). When all 5 MLVA loci were used, the correlation of pairwise MLVA distance to pairwise SNP distance was $0\mathord{\cdot}77$. However, the full set of loci did not produce the highest correlation to SNP counts. There was large variation of correlations between the loci, with locus 2 in particular showing a weak correlation of $0\mathord{\cdot}07$. The highest correlation of MLVA to SNP counts was for the combination of loci 1 and 5, at $0\mathord{\cdot}83$.

\newpage

\newpage

\newpage
\begin{figure}[h]
    \centering
    \includegraphics[scale=0.8]{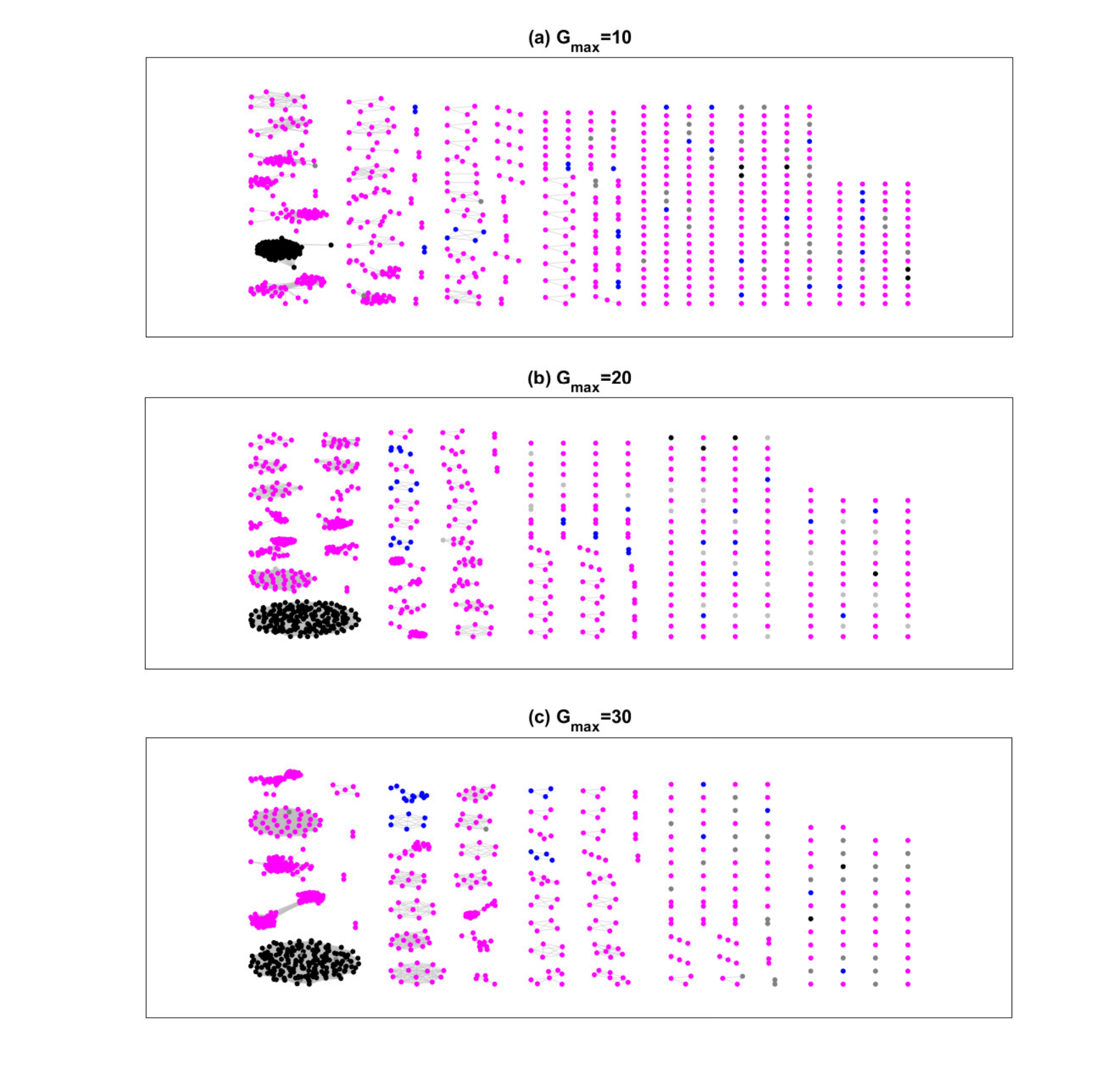}
    \caption{The undirected thresholded SNP sub-network. SNP distance threshold $G_{max}=10, 20$ and $30$ in (a), (b) and (c) respectively. Corresponding network properties are shown in Table \ref{fig:undirected_network_properties}. Components highlighted in black, magenta, and blue corresponds to MLVA components identified in Fig.~\ref{fig:undirected_net_MLVA}. }
    \label{fig:Fig6_sensi}
\end{figure}

\newpage
\begin{figure}[h]
    \centering
    \includegraphics[scale=0.8]{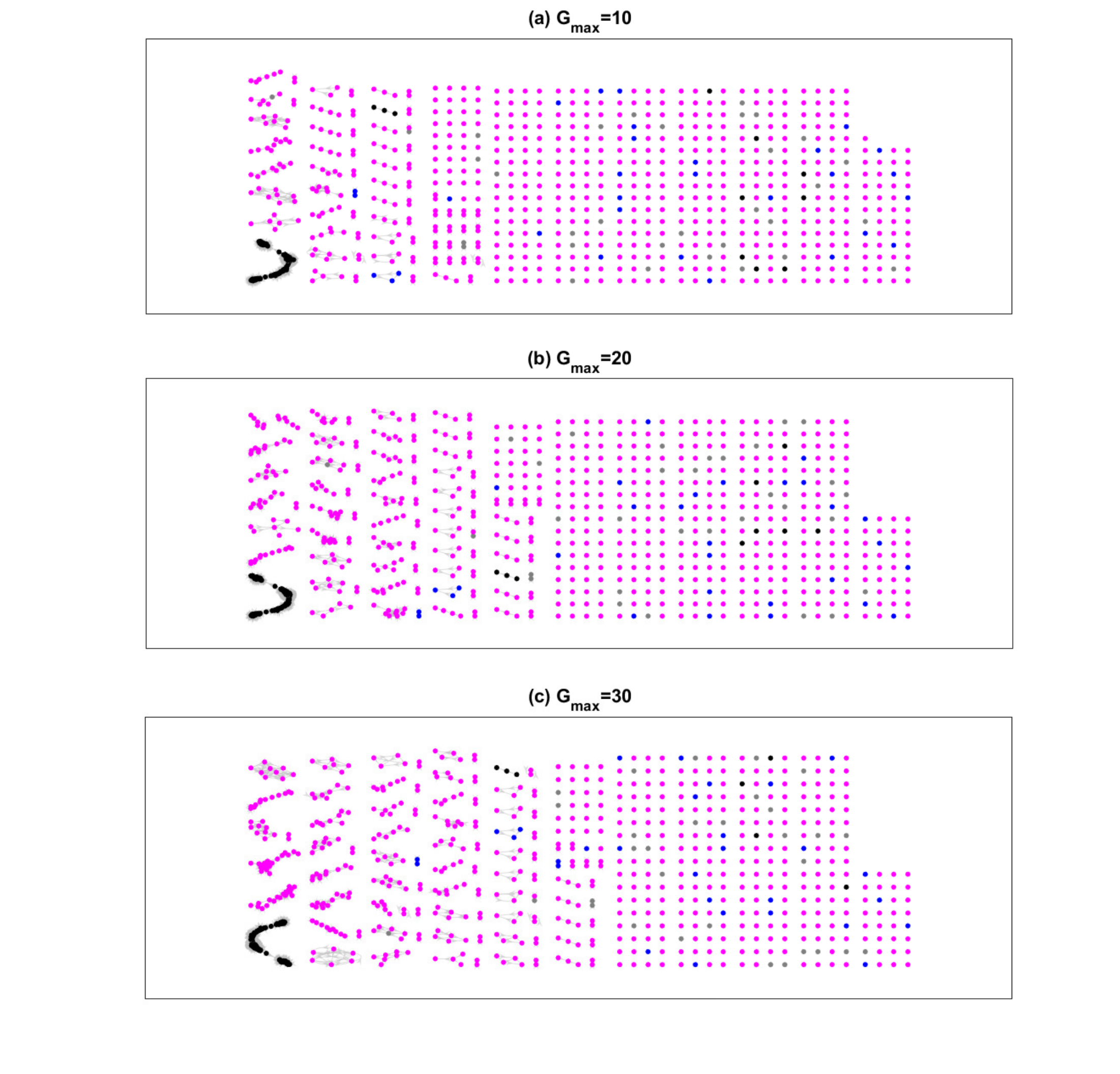}
    \caption{The directed SNP sub-network. SNP distance threshold $G_{max}=10, 20$ and $30$ in (a), (b) and (c) respectively. Corresponding network properties are shown in Table \ref{fig:directed_network_properties}. Time window $T_{max} = 30$ days. Components highlighted in black, magenta, and blue corresponds to MLVA components identified in Fig.~\ref{fig:undirected_net_MLVA}.}
    \label{fig:Fig7_sensi}
\end{figure}

\newpage
\begin{figure}[h]
    \centering
    \includegraphics[scale=0.8]{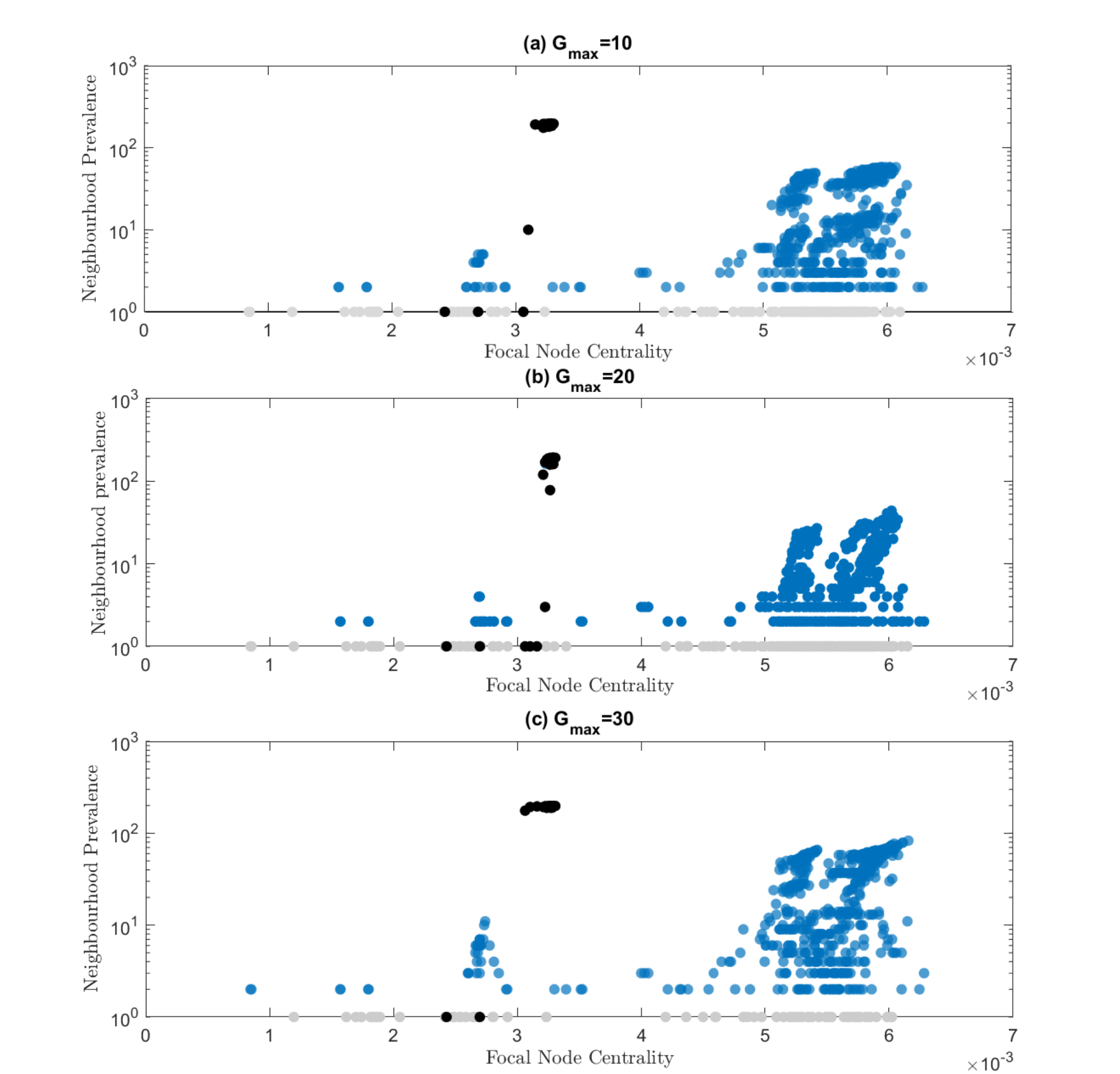}
    \caption{Centrality-prevalence space for the undirected SNP network where each node represents an isolate, mapping the centrality of profiles (x-axis) and their neighbourhood prevalence (y-axis). Neighbourhood threshold $G_{max}=10, 20$ and $30$ in (a), (b) and (c) respectively. Disconnected nodes are shown in grey. Isolates highlighted in black belong to the component shown in black in Fig.~\ref{fig:undirected_net_MLVA}. The nodes shown in blue belong to non-outbreak components. Note that since each isolate is unique, the neighbourhood prevalence here is interpreted as the neighbourhood size.}
    \label{fig:Fig8_sensi}
\end{figure}

\definecolor{lightGreen}{rgb}{0.412, 0.631, 0.357}
\definecolor{darkGreen}{rgb}{0.102, 0.435, 0.02}
\definecolor{peach}{rgb}{0.918, 0.506, 0.412}
\definecolor{lightOrange}{rgb}{0.965, 0.827, 0.616}
\definecolor{lilac}{rgb}{0.796, 0.58, 0.89}
\definecolor{khaki}{rgb}{0.533, 0.635, 0.278}
\definecolor{purple}{rgb}{0.38, 0.278, 0.635}
\definecolor{aqua}{rgb}{0.318, 0.58, 0.588}
\definecolor{desatRed}{rgb}{0.749, 0.376, 0.404}
\definecolor{yellow}{rgb}{0.965, 0.871, 0.298}
\definecolor{darkPink}{rgb}{0.89, 0.075, 0.541}
\definecolor{orange}{rgb}{0.965, 0.608, 0.298}
\definecolor{pink}{rgb}{0.851, 0.075, 0.89}
\definecolor{cyan}{rgb}{0.075, 0.89, 0.773}
\definecolor{red}{rgb}{0.89, 0.075, 0.137}
\definecolor{darkBlue}{rgb}{0.075, 0.165, 0.89}
\definecolor{purple}{rgb}{0.553, 0.075, 0.89}
\definecolor{lime}{rgb}{0.694, 0.89, 0.075}
\definecolor{lightBlue}{rgb}{0.298, 0.655, 0.965}
\definecolor{lightGray}{rgb}{0.647, 0.675, 0.839}
\definecolor{brightOrange}{rgb}{0.925, 0.517, 0.078}
\definecolor{desatPurple}{rgb}{0.612, 0.392, 0.58}
\definecolor{darkGray}{rgb}{0.165, 0.176, 0.31}

\newpage
\begin{table}[h]
    \centering
    \renewcommand{\arraystretch}{1.18}
    \begin{tabular}{|p{2cm}|c|c|}
    \hline
         Colour & Name & Number of isolates \\
    \hline
    \hline
    \multicolumn{3}{|c|}{Network Components (\ref{fig:phylo}A)}\\
    \hline
          \cellcolor{gray} & Singletons \& small components  & 391 \\
          \cellcolor{black} & Component 1 & 196 \\
          \cellcolor{lightGreen}& Component 2 & 65 \\
          \cellcolor{darkGreen} & Component 3 & 60 \\
          \cellcolor{peach} & Component 4 & 51 \\
          \cellcolor{lightOrange} & Component 5 & 37 \\
          \cellcolor{lilac} & Component 6 & 25 \\
          \cellcolor{khaki} & Component 7 & 23 \\
          \cellcolor{purple} & Component 8 & 20 \\
          \cellcolor{aqua} & Component 9 & 15 \\
          \cellcolor{desatRed} & Component 10 & 14 \\
    \hline
    \multicolumn{3}{|c|}{SNP Clusters (\ref{fig:phylo}A)}\\
    \hline
          \cellcolor{gray} & Singletons \& small clusters & 582\\
          \cellcolor{yellow} & SalEnt-16-0001 & 19 \\
          \cellcolor{darkPink} & SalEnt-16-0005 & 10 \\ 
          \cellcolor{orange} & SalEnt-16-0006 & 20 \\
          \cellcolor{pink} & SalEnt-16-0016 & 7 \\
          \cellcolor{cyan} & SalEnt-16-0031 & 13 \\
          \cellcolor{red} & SalEnt-17-0002 & 12 \\
          \cellcolor{darkBlue} & SalEnt-17-0007 & 14 \\
          \cellcolor{purple} & SalEnt-18-0027 & 11 \\
          \cellcolor{lime} & SalEnt-18-0029 & 18 \\
          \cellcolor{lightBlue} & SalEnt-18-0030 & 191 \\
    \hline
    \multicolumn{3}{|c|}{MLVA Profiles (\ref{fig:phylo}B)}\\
    \hline
         \cellcolor{gray} & No profile & 7\\
         \cellcolor{gray} & 2-10-NA-5-1 & 1\\
         \cellcolor{brightOrange} & 2-10-8-5-1 & 187\\
         \cellcolor{desatRed} & 2-10-8-6-1 & 1\\
         \cellcolor{black} & 2-10-9-5-1 & 1\\
         \cellcolor{desatPurple} & 2-11-8-5-1 & 1\\
         \cellcolor{lightGray} & 2-12-8-5-1 & 1\\
         \cellcolor{darkGray} & 2-13-8-5-1 & 2\\
    \hline     
    \end{tabular}
    \vspace{2mm}
    \caption{Colours used in Fig.~\ref{fig:phylo}.}
    \label{fig:colour}
\end{table}

\newpage
\begin{table}[h]
    \centering
    \renewcommand{\arraystretch}{1.18}
    \begin{tabular}{|c|c|c|c|c|c|}
    \hline
     $G_{max}$  & Edges, M & Singletons & Components  &  Components    & Largest  \\
          &  &  &  (incl. singletons) &  ($N \geq 3$)   &  component \\
    \hline
    10 & 18,392 & 277   & 351             & 36 & 194             \\
    \hline
    20 & 22,742 & 162   & 229             & 42 & 196              \\
    \hline
    30 & 26,238 & 87    & 148             & 44 & 198               \\
    \hline
    \end{tabular}
    \vspace{2mm}
    \caption{Properties of the undirected network in Fig.~\ref{fig:Fig6_sensi}.}
    \label{fig:undirected_network_properties}
\end{table}

\newpage
\begin{table}[h]
    \centering
    \renewcommand{\arraystretch}{1.18}
    \begin{tabular}{|c|c|c|c|c|c|c|c|}
    \hline
     $G_{max}$  & Edges, M & Singletons & Components  &   Components    & Largest  & Longest  & Average directed  \\
            &  &  & (incl. singletons) &   ($N \geq 3$)   &  component & path &  path length \\
    \hline
     10 & 4,057 & 496       & 567    & 30  & 188 & 17 & 4.55         \\
    \hline
     20 & 4,252 & 400       & 483    & 41  & 191 & 17 & 4.54         \\
    \hline
     30 & 4,678 & 330       & 420    & 48  & 193 & 17 & 4.48         \\
    \hline
    \end{tabular}
    \vspace{2mm}
    \caption{Properties of the directed network in Fig.~\ref{fig:Fig7_sensi}.}
    \label{fig:directed_network_properties}
\end{table}
\end{document}